 \documentclass[preprint,showpacs,preprintnumbers,amsmath,amssymb]{revtex4}

\usepackage{graphicx}
\usepackage{epsfig}
\usepackage{epsf}

\usepackage{dcolumn}
\usepackage{bm}

\newcommand{\beq}{\begin{equation}}
\newcommand{\eeq}{\end{equation}}
\newcommand{\beqn}{\begin{eqnarray}}
\newcommand{\eeqn}{\end{eqnarray}}
\newcommand{\bea}[1]{\beq\begin{array}{#1}}
\newcommand{\eea}{\end{array}\eeq}

\def\lsim{\raise0.3ex\hbox{$<$\kern-0.75em\raise-1.1ex\hbox{$\sim$}}}
\def\gsim{\raise0.3ex\hbox{$>$\kern-0.75em\raise-1.1ex\hbox{$\sim$}}}

\begin{document}

\title{Spatial string tension in $N_f=2$ lattice QCD at finite temperature} 

\author{V.G.~Bornyakov,$^{1,2}$ E.V.~Luschevskaya$^{2}$ }
\affiliation{
$^1$Institute for High Energy Physics IHEP, 142281 Protvino, Russia\\
$^2$Institute of Theoretical and
Experimental Physics ITEP, 117259 Moscow, Russia \\
\vspace*{-0.75cm}
}

\begin{abstract}
The spatial string tension across a crossover from the low temperature phase to the
high temperature phase is computed in QCD with two flavors of non-perturbatively improved Wilson fermions at small lattice spacing  $a \sim 0.12$~fm. 
We find that in the low temperature phase spatial string tension agrees 
well with zero temperature string tension. Furthermore, it does not show increasing for temperatures up to  $T = 1.36 T_{pc}$, the highest temperature considered.
Our results agree with some theoretical predictions.
\end{abstract}


\maketitle

\section{Introduction}
\label{one}

In view of running and future heavy-ion collisions experiments  
determination of the (pseudo)critical
temperature $T_{pc}$ as well as studies of other features of the  finite temperature
transition in full QCD with  $N_f=2$ and $N_f=2+1$ flavors of dynamical quarks
are subjects of many recent lattice investigations.  
Three groups employed improved staggered fermions and an improved gauge field
action~\cite{Aoki:2006br,Cheng:2006qk,Bernard:2006nj}. Simulations with  staggered fermions, although computationally less demanding, are made with the use of the fourth root of the fermion
determinant.  This introduces theoretical uncertainty. Thus studies with the Wilson-type  quark 
actions free of such problem are necessary.
The CP-PACS \cite{AliKhan:2000iz},  WHOT-QCD \cite{WHOT-QCD} and DIK \cite{Bornyakov:2005dt} collaborations used improved Wilson fermions with an improved or unimproved 
gauge field action.  In this paper we are using the lattice configurations generated by DIK
collaboration.

Recent results~\cite{Aoki:2006br,Bornyakov:2005dt} show importance of the
finite lattice spacing effects in the studies of the finite temperature transition 
in QCD.
The DIK collaboration simulations were done on
lattices with large $N_t$, i.e. the transition was studied
at small lattice spacing, close to the continuum limit. 
For this reason we consider it important to study in more details various features of 
the finite temperature QCD using gauge field configurations accumulated
by DIK collaboration. 
In this letter we present results for the spatial string tension $\sigma_s$. 

In the presence of dynamical quarks the flux tube formed between static
quark-antiquark pairs breaks at large distances. 
In contrast the so called spatial Wilson loop, i.e. the Wilson loop with both 
directions being spatial is found to show non-vanishing spatial string tension 
even at $T>T_{pc}$. This is called the spatial confinement.
$\sigma_s$ is one of the fundamental quantities to characterize the perturbative and non-perturbative properties of the hot QCD medium. It provides useful information about  
the effective 3D model, which can be obtained by integrating out heavy modes.
It is believed that non-perturbative contributions to observables, which 
at high temperature are dominated by contributions of magnetic modes, are 
flavor independent apart from indirect dependence through the flavor dependence
of the running coupling. $\sigma_s$  can be used to test this
aspect of the dimensional reduction approach to QCD. 

Although quarks are expected to decouple from the
spatial observables for $T \gg T_{pc}$ due to dimensional reduction
and thus do not affect $\sigma_s$ in the high temperature limit,
it is not obvious whether the same is true near $T_{pc}$.
Thus lattice computations are also useful to check theoretical predictions for this range
of temperatures. 
\begin{table*}
\begin{center}
\mbox{
\begin{tabular}{|c|c|c|c|c|c|c|}\hline
$\beta$ &$\kappa$ & $a/r_0$ & $m_\pi r_0 $ & $T/T_{pc}$ & $\sigma_s a^2$ & $\sqrt{\sigma} r_0$ \\  \hline
  5.2   & 0.1330  & 0.2854  &    2.89      & 0.78       & 0.110 (3)  & 1.16\\
  5.25  & 0.1335  & 0.2425  &    2.98      & 0.92       & 0.085 (2)  &1.16 \\
  5.25  & 0.1340  & 0.2272  &    2.72      & 0.99       & 0.069 (4)  &1.16 \\
  5.25  & 0.1341  & 0.2242  &    2.66      & 1.01       & 0.063 (3)  &1.15 \\
  5.25  & 0.1342  & 0.2213  &    2.60      & 1.03       & 0.063 (2)  &1.15 \\
  5.2   & 0.1348  & 0.2192  &    2.03      & 1.07       & 0.058 (2)  &1.14\\
  5.2   & 0.1360  & 0.1846  &    1.06      & 1.36       & 0.040 (2)  &1.13\\
\hline
\end{tabular}
}
\end{center}
\caption{The parameters of lattice configurations used in computations of $\sigma_s $.  $\sqrt{\sigma} r_0$ is the physical string tension dependent on $m_\pi r_0 $, respective values are taken from \cite{wuppertal}}
\label{tab:configs}
\end{table*}

\section{Details of computations}
$\sigma_s$  was recently computed by 
RBC-Bielefeld \cite{RBC-Bielefeld,Liddle:2007uy} and WHOT-QCD \cite{WHOT-QCD} 
collaborations.  
Our lattice action is different from that used in \cite{RBC-Bielefeld,Liddle:2007uy} and
similar to action used in \cite{WHOT-QCD}. The important difference from 
\cite{WHOT-QCD} is that we are working at two times smaller lattice spacing, i.e. much closer
to the continuum limit. 
We perform computations on $N_s^3\,N_t=16^3\,8$
lattices while in \cite{WHOT-QCD} $N_t=4$ lattices were used.
The configurations were generated with the usual Wilson gauge field action 
and the non-perturbatively $O(a)$ improved Wilson fermionic action $S_F$:
\begin{equation}
S_F = S^{(0)}_F - \frac{\rm i}{2} \kappa\, g\,
c_{sw} a^5
\sum_s \bar{\psi}(s)\sigma_{\mu\nu}F_{\mu\nu}(s)\psi(s),
\label{action}
\end{equation}
where $S^{(0)}_F$ is the original Wilson fermionic action, $g$ is the gauge
coupling and $F_{\mu\nu}(x)$ is the field strength tensor. The
clover coefficient $c_{sw}$ is determined non-perturbatively \cite{Jansen:1998mx}. 
We use configurations generated at two values of the lattice coupling $\beta=6/g^2$: 
$\beta = 5.2$ and $\beta = 5.25$, for temperatures between $0.78 ~T_{pc}$ and 
$1.36 ~T_{pc}$.
To fix the physical scale and the quark mass results
obtained with this action by QCDSF and UKQCD
collaborations~\cite{Booth:2001qp,Allton:2002sk,Gockeler:2006vi} were used.
The  scale is fixed by the phenomenological length $r_0$ \cite{sommer} and the quark 
mass is fixed by $m_{\pi}r_0$ value. The parameters of our lattices are presented in 
the TABLE~\ref{tab:configs}. Note, that we used for $T_{pc}$  the value depending on $m_{\pi}r_0$
which was obtained using interpolation of DIK results as function of  $m_{\pi}r_0$ \cite{Bornyakov:2007zu}. For this reason values for $T/T_{pc}$ in TABLE~\ref{tab:configs} are slightly different from those in  \cite{Bornyakov:2005dt}.

We evaluate $\sigma_s$ following computational procedure used in 
\cite{Bali:1993tz,Boyd:1996bx} where pure gauge theories were investigated. 
Let us consider $z$ direction as the 'time' direction of the spatial Wilson
loops.
The temperature dependent spatial static potential $V_{s}(R)$ 
is then defined as 
\begin{equation}
aV_{s}(R)  = \lim_{Z \to \infty} \log \frac{W(R,Z)}{W(R,Z+a)}\,,
\label{potential1}
\end{equation}
where $W(R,Z)$ are Wilson loops of size $R/a \times Z/a$  in lattice units.
We assume an  ansatz for $V_{s}(R)$ including linear and selfenergy terms: 
\begin{equation}
V_{s}(R)  = V_0 +\sigma R\,. \label{poten}
\end{equation}
The ambiguous Coulomb term is not included and instead we make our fits at large distances
$R/a \ge 5$ to avoid such term contribution.  
To increase the overlap with the respective ground state the APE 
smearing \cite{Albanese:1987ds} of the links orthogonal to this direction 
was used. 
Furthermore, the hypercubic blocking (HYP) procedure \cite{Hasenfratz:2001hp} 
was applied to links in $z$-direction to 
decrease the selfenergy $V_0$ and thus to decrease the statistical errors.    
We applied three steps of HYP. It occurred that the first step produced the 
maximal effect  decreasing  statistical errors of  the potential $V_{s}(R)$  by factor
about 5. Application of the second and third steps of HYP were much less effective but still useful, decreasing statistical errors by about 20\%. 
\begin{figure}[phtb]
\centerline{\includegraphics[angle=0,scale=0.90,clip=false]{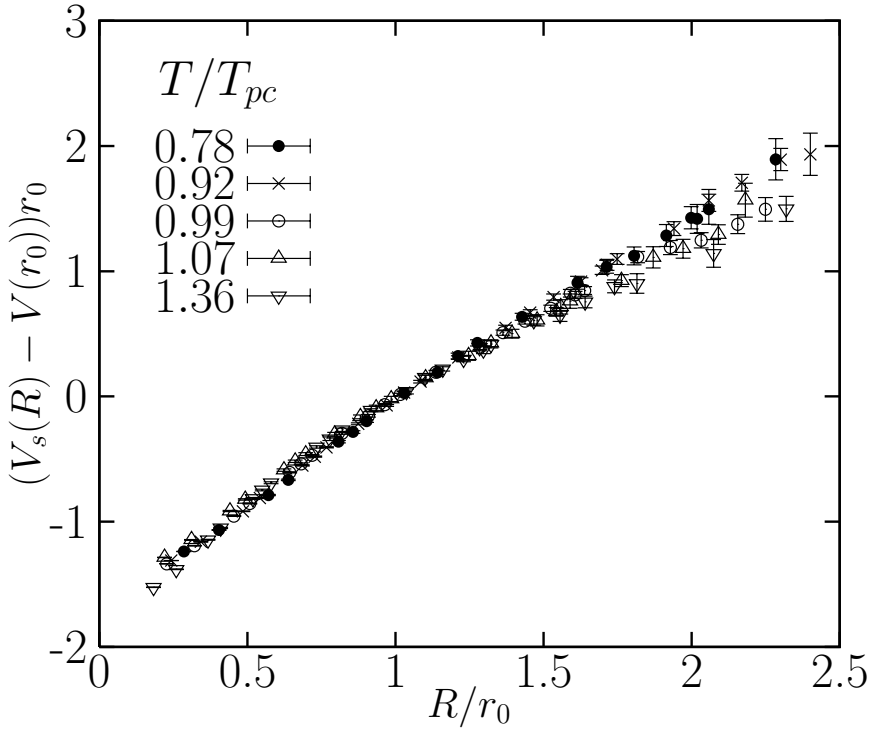}}
\caption{Spatial static potential $V_{s}(R)$.}
\label{fig:potential}
\end{figure}
The limit $Z \to \infty$ in (\ref{potential1}) was approximated by taking 
the spatial Wilson loops with extension  $Z/a= 4$ or 5 to evaluate $V_{s}(R)$. 

\section{Results}
\begin{figure}[!htb]
\centerline{\includegraphics[angle=0,scale=0.90,clip=false]{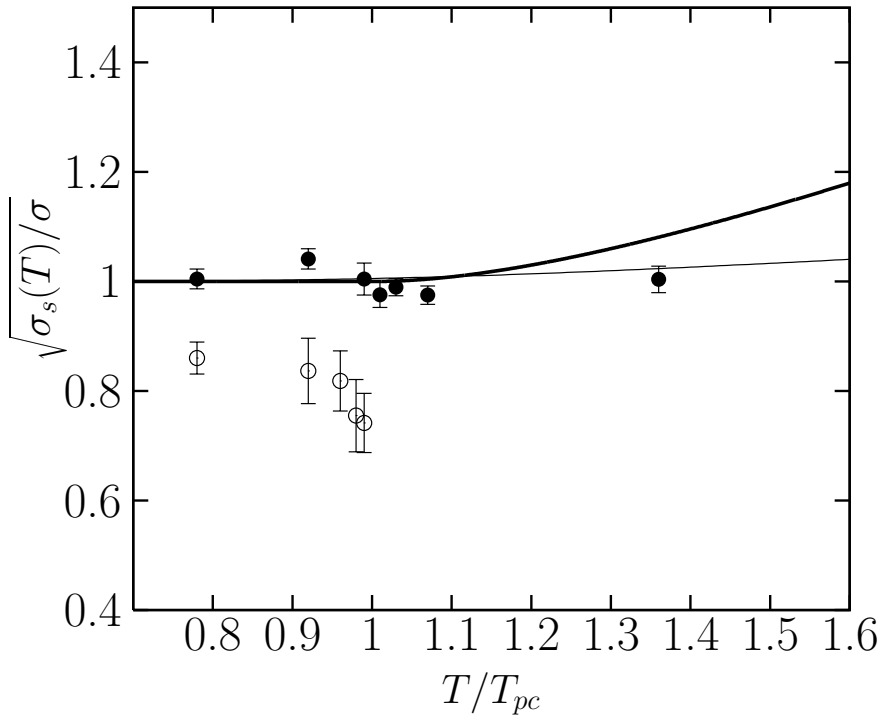}}
\caption{Square root of the ratio $\sigma_s(T)/\sigma$ (full circles) as function of 
temperature. The values of $\sigma$ are shown in Table I, see explanation in the text. 
Respective data for the finite temperature physical string tension $\sigma(T)$ (empty circles) computed by DIK \cite{Bornyakov:2004ii} is shown for comparison. The curves are theoretical predictions from Ref.~\cite{Agasian:2003yw} (thin curve) and Ref.~\cite{VIZ} (thick curve).}
\label{fig:string_tension_1}
\end{figure}

In Fig.~\ref{fig:potential} we show the potential $V_{s}(R)$ with $V_{s}(r_0)$ subtracted as function of $R/r_0$ for different temperatures.
One can see that potentials for all temperatures agrees with each other quite
well, i.e. there is no strong temperature dependence in the temperature
interval considered. 
Respectively the spatial string tension is also virtually independent of 
temperature as can be seen from Fig.~\ref{fig:string_tension_1}.  
In this figure we present our results for $\sigma_s(T)$ as a dimensionless ratio 
$\sqrt{\sigma_s(T)/\sigma}$, where $\sigma$ is the physical string tension at zero temperature \footnote{It is known that at zero temperature the physical string tension $\sigma$ decreases
with decreasing of the quark mass \cite{wuppertal}. We took this into account using for $\sigma$ results of Ref.~\cite{wuppertal}  for $m_{\pi} r_0$ values close to our respective values as shown in Table~\ref{tab:configs}.}.  We also depicted in this figure DIK results for the finite temperature physical string tension  $\sigma(T)$ to emphasize drastic difference in the temperature dependence of  $\sigma(T)$ and $\sigma_s(T)$. 

Our results for $\sigma_s(T)$ allow to conclude that for temperatures below $T_{pc}$ 
$\sigma_s(T)$ is constant and equal to $\sigma$. This does not come as a surprise since such equality was observed in the quenched theory long ago \cite{Bali:1993tz,Boyd:1996bx}.
Unexpectedly the only previous  result for $T<T_{pc}$ in unquenched lattice QCD~\cite{WHOT-QCD}   contradicts our result. In  Ref.~\cite{WHOT-QCD} $\sqrt{\sigma_s(T)}$  for $T<T_{pc}$ is roughly $3 T_{pc}$ i.e. essentially higher then $sqrt{\sigma}$. The reason for such confusing result might be that in Ref.~\cite{WHOT-QCD} different method to extract the string tension $\sigma_s$  was used.
At $T>T_c$ our results are also different from those of Ref.~\cite{WHOT-QCD} although agree qualitatively with results of Ref.~\cite{Liddle:2007uy}. 

Let us now compare our results with two theoretical predictions. In Ref.~\cite{Agasian:2003yw}
the following expression for $\sigma_s(T)$ at $T<2T_{pc}$  was derived within the field 
correlators  approach \cite{DiGiacomo:2000va}:
\begin{equation}
 \frac{\sigma_s(T)}{\sigma} = \frac{\sinh(M/T) + M/T}{\cosh(M/T)+1}\,, 
\label{agasian}
\end{equation}
where $M$ is the inverse magnetic correlation length and is taken in Ref.~\cite{Agasian:2003yw}
equal to 1.5~GeV. Plotting square root of expression (\ref{agasian}) in Fig.~2 we took $M/T_c=7.5$
thus taking $T_c=200$ MeV. Note that in the considered temperature range the precise value of $M/T_c$ is not important.  When $M/T_c >> 1$ equation (\ref{agasian}) suggests very slow increase of the spatial string tension with temperature in nice agreement with our numerical results. 
Recently prediction for $\sigma_s(T)$ was derived via AdS/QCD approach \cite{VIZ}:
\begin{equation}
\frac{\sigma_s(T)}{\sigma} = \left ( \frac{T}{T_c} \right )^2 \exp{\left[\left(\frac{T}{T_c}\right)^2-1\right]}\,,\quad T \ge T_c\,,
\label{viz}
\end{equation}
and $\sigma_s(T)=\sigma$ for $T \le T_c$. This expression also provide qualitatively good description of our data, as can be seen from Fig.~2.

We shall note that in our computations we do not keep the pion mass fixed, i.e. the computations are not made along the constant physics line. This may introduce sizable systematic effects if $\sigma_s/\sigma$ has strong dependence on the quark mass. In Ref.~\cite{WHOT-QCD}
$\sigma_s$ was computed for two values of $m_\pi/m_\rho$, 0.65 and 0.85. The mild dependence on the quark mass  was observed. These results need to be confirmed. In any case our conclusion about independence of  $\sigma_s$ on the temperature below transition is intact since $m_\pi r_0$ varies by only about 10\% when temperature increases from $T/T_{pc}=0.78$ to $T/T_{pc}=1.01$, as can be seen from the Table \ref{tab:configs}.

Let us summarize our main results.  
We presented first numerical evidence in unquenched lattice QCD that for temperatures below $T_{pc}$ $\sigma_s(T)$ is constant and equal to $\sigma$, the physical string tension at $T=0$.  
This is in agreement with quenched lattice QCD results and theoretical predictions.
The theoretical predictions eq.(\ref{agasian}) \cite{Agasian:2003yw}  and eq(\ref{viz}) \cite{VIZ} for $\sigma_s(T)$ at temperatures $T_{pc} < T < 1.4 T_{pc}$ where all known lattice results clearly deviate from the expected high temperature behavior 
\begin{equation}
\sqrt{\sigma_s(T)} = c g^2(T) T
\end{equation}
are in good agreement with our results with better agreement for eq.(\ref{agasian}).

\begin{acknowledgments}
We wish to thank N.O.~Agasian, M.I.~Polikarpov and Yu.A.~Simonov for 
useful discussions. This work was completed due to finacial support by the 
Russian Federal Agency of Atomic Energy.
It is also partially supported by grants RFBR 05-02-16306, 
RFBR 07-02-00237,  RFBR-DFG 06-02-04010. 
\end{acknowledgments}

\end{document}